# Focusing of Spin Polarization in Semiconductors by Inhomogeneous Doping

Yuriy V. Pershin  and  Vladimir Privman

*Center for Quantum Device Technology, Clarkson University,*
*Potsdam, New York 13699-5720, USA*

**Abstract:**  We study the evolution and distribution of non-equilibrium electron spin polarization in n-type semiconductors within the two-component drift-diffusion model in an applied electric field. Propagation of spin-polarized electrons through a boundary between two semiconductor regions with different doping levels is considered. We assume that inhomogeneous spin polarization is created locally and driven through the boundary by the electric field. The electric field distribution and spin polarization distribution are calculated numerically. We show that an initially created narrow region of spin polarization can be further compressed and amplified near the boundary. Since the boundary involves variation of doping but no real interface between two semiconductor materials, no significant spin-polarization loss is expected. The proposed mechanism will be therefore useful in designing new spintronic devices.

---

**Introduction:** Theoretical and experimental investigations of spin-related effects in semiconductors have received much attention recently [1-5] owing to the proposals for devices based on the manipulation of electron spin [6-16]. Operation of a spintronic device requires efficient spin injection into a semiconductor, spin manipulation, control and transport, and also spin detection. Once injected into a semiconductor, electrons experience spin-dependent interactions with the environment, which cause relaxation. Due to diffusion, spin polarization spreads over the sample and spin polarization density decreases as well. For effective spin manipulation and detection, it is desirable to have high spin polarization densities—the problem addressed in the present work.

Non-equilibrium spin polarization can be introduced into a semiconductor in various ways. Experimentally, it is realizable at the interface between semiconductor and ferromagnetic metal or magnetic semiconductor [17,18], or by using optical pumping techniques [19-21]. Another possible approach is to use hyperfine interaction of electrons with polarized nuclei [22]. In the latter scheme, the nuclear spins should be first polarized by optical pumping, or by spin-polarized current [7,23,24], or (anti)ferromagnetically ordered at ultra-low temperatures [25]. In this work, we consider spin-polarization created locally in the bulk of semiconductor, for example, by using ferromagnetic-metal scanning tunneling microscopy (STM) tips [26,27], or by optical pumping techniques [19-21].

Figure 1 shows the system under investigation. Localized electron spin polarization is created by a source in the semiconductor region with the doping density $N_1$. Under influence of the applied electric field, the spin-polarized carriers drift through the boundary between the two semiconductor regions, i.e., from the region with the doping density $N_1$ to the region with the doping density $N_2$. It is assumed that $N_2 \geq N_1$. In our calculations, we use the two-component drift-diffusion model [28,29], and we take into account the effects of the charge accumulation/redistribution near the boundary. The latter effect is analogous to the depletion region formation in a p-n junction [30], and it introduces coordinate-dependent electric field in the equation for the spin polarization density. We solve the resulting differential equations for the electric field and spin polarization density numerically. Two types of spin-polarization source are considered: instantaneous source and continuous one. The main result of our calculations is that the spin polarization density can be condensed and amplified near the boundary. The system of interest does not require real interface between semiconductor materials. We only assume variation in the doping level. Therefore, additional electron spin-polarization losses at interfaces [31] can be avoided. This result will be useful in designing new spintronic devices.

**Model:** Our theoretical investigation is based on the two-component drift-diffusion model (see, e.g., [29]). In our case, the system is described by the following set of equations:

$$e\frac{\partial n_{\uparrow(\downarrow)}}{\partial t} = \text{div }\vec{j}_{\uparrow(\downarrow)} + \frac{e}{2\tau_{sf}}\left(n_{\downarrow(\uparrow)} - n_{\uparrow(\downarrow)}\right) + S_{\uparrow(\downarrow)}(\vec{r},t) \quad , \qquad (1)$$

$$\vec{j}_{\uparrow(\downarrow)} = \sigma_{\uparrow(\downarrow)}\vec{E} + eD\nabla n_{\uparrow(\downarrow)} \quad , \qquad (2)$$

and

$$\sigma_{\uparrow(\downarrow)} = en_{\uparrow(\downarrow)}\mu \quad , \qquad (3)$$

where $e$ is the electron charge, $n_{\uparrow(\downarrow)}$ is the density of spin-up (spin-down) electrons, $\vec{j}_{\uparrow(\downarrow)}$ is the current density, $\tau_{sf}$ is the spin relaxation time, $S_{\uparrow(\downarrow)}(\vec{r},t)$ describes the source of the spin polarization, $\sigma_{\uparrow(\downarrow)}$ is the conductivity, and $\mu$ is the mobility, connected with the diffusion coefficient $D$ via the Einstein relation $\mu = De/(k_B T)$, and defined via $\vec{v}_{drift} = \mu\vec{E}$.

Equation (1) is the usual continuity relation that takes into account the spin relaxation and the source of the spin polarization, Eq. (2) is the expression for the current which includes the drift and the diffusion contributions, and Eq. (3) is the expression for the conductivity. It is assumed that the diffusion coefficient $D$ and the spin relaxation time $\tau_{sf}$ are equal for spin-up and spin-down electrons.

To separate the equations for charge and spin degrees of freedom, we introduce the charge density $n = n_\uparrow + n_\downarrow$ and the spin polarization density $P = n_\uparrow - n_\downarrow$. Behavior of the charge density is given by the equations

$$\vec{j}_0 = en\mu\vec{E} + eD\nabla n \qquad (4)$$

and

$$\text{div }\vec{E} = \frac{e}{\varepsilon\varepsilon_0}(N_i - n) \quad . \qquad (5)$$

Here $\vec{j}_0$ is the current flowing through the sample, $\varepsilon_0$ is the permittivity of the free space, and $\varepsilon$ is the dielectric constant. We assume that at room temperature the density of the ionized donors $N_i$ is equal to the donor density ($N_i = N_1$ for $x < 0$ and $N_i = N_2$ for $x > 0$), i.e., all the donors are ionized. Equation (4) was obtained from Eqs. (1)-(3) neglecting the term which describes the source of the spin polarization, because we have assumed that charge equilibration processes at the point of injection do not significantly influence the electric field profile and thus the propagation of the spin polarization through the boundary. Relation (5) is the Poisson equation. Combining Eq. (4) and Eq. (5), we obtain the equation describing the electric field distribution,

$$\frac{\partial^2 E}{\partial x^2} + \frac{e}{kT} E \frac{\partial E}{\partial x} - \frac{e^2 N_i}{kT\varepsilon\varepsilon_0} E = -\frac{j_0}{\varepsilon\varepsilon_0 D} + \frac{e}{\varepsilon\varepsilon_0} \nabla N_i \quad , \quad (6)$$

where $E$ in $x$-direction.

From the set of relations (1)-(3), we obtain an equation for the spin polarization density,

$$\frac{\partial P}{\partial t} = D\Delta P + D\frac{e\vec{E}}{k_B T}\nabla P + D\frac{e\nabla \vec{E}}{k_B T}P - \frac{P}{\tau_{sf}} + F(\vec{r},t) \quad . \quad (7)$$

Here $F(\vec{r},t) = [S_\uparrow(\vec{r},t) - S_\downarrow(\vec{r},t)]/e$ represents a spin polarization density created by the external source. The spin polarization density is coupled to the charge density through the electric field. Thus, our numerical calculation involves two steps: first, the electric field profile is found as the solution of Eq. (6) and, second, Eq. (7) is solved for the spin polarization density.

**Results and discussion:** To proceed with solution of Eq. (6), let us introduce the dimensionless variables as $\tilde{E} = E/E_0$ and $\tilde{x} = x/x_0$, where $E_0 = j_0 kT/(DN_1 e^2)$ and $x_0^{-2} = e^2 N_1/(kT\varepsilon\varepsilon_0)$. Equation (6) can be rewritten as

$$\frac{\partial^2 \tilde{E}}{\partial \tilde{x}^2} + \alpha\, \tilde{E}\, \frac{\partial \tilde{E}}{\partial \tilde{x}} - \frac{N_i}{N_1}\tilde{E} + 1 - \beta\delta(\tilde{x}) = 0 \quad , \quad (8)$$

where $\alpha = x_0 E_0 e/kT$ and $\beta = x_0 e(N_2 - N_1)/(E_0 \varepsilon\varepsilon_0)$. The estimation of the dimensionless constant $\alpha$ and $x_0$ for $E = 10^3$ V/cm and $N_1 = 10^{15}$ cm$^{-3}$, gives $x_0 = 1.37\times 10^{-7}$ m and $\alpha = 0.52$. The solution of Eq. (8) was found numerically. Figure 2 shows the electric field profile near the boundary for selected values of the parameters.

It is also convenient to rewrite Eq. (7) in the dimensionless form. With the dimensionless variables selected as follows, $\tau = t/\tau_{sf}$, $X = x/x_1$, with $x_1 = \sqrt{D\tau_{sf}}$, we get

$$\frac{\partial p}{\partial \tau} = \frac{\partial^2 p}{\partial X^2} + \sqrt{D\tau_{sf}}\frac{e}{k_B T}\left(E\frac{\partial p}{\partial X} + p\frac{\partial E}{\partial X}\right) - p + g \quad . \quad (9)$$

The source function $g$ and dimensionless spin polarization density are defined for instantaneous and continuous sources in different ways. Assuming an instantaneous source, we take the function $F$ in Eq. (7) as $F = F_0 f(x)\delta(t)$ with $f(x)$ normalized to 1; then, $g = x_1 f(X)\delta(\tau)$ and $p = Px_1/F_0$.

Here $F_0$ measures the spin polarization density created at the initial moment of time. The continuous source $F = G_0 \delta(x)$ is described by the constant $G_0$, which measures the spin polarization density created per unit time. The dimensionless polarization density in this case is defined as follows, $p = P x_1 / F_0 \tau_{sf}$, $g = \delta(X)$.

Evolution of the spin polarization density created at $t = 0$ is shown in Fig. 3. We have selected the profile of initial spin polarization in the Gaussian form and solved Eq. (9) using an iterative scheme. Under influence of the electric field, the spin polarization density profile moves towards the boundary. Its width increases due to diffusion process, and the amplitude decreases due to the combined action of the different spin relaxation mechanisms and diffusion processes. As the spin polarization density profile approaches the boundary, its velocity increases. It reaches the maximum at the boundary, where the electric field is maximal. In the region with higher donor density $N_2$, the electric field is lower, and the velocity of the spin polarization profile decreases. As a result, the spin polarization gathers in a narrow spatial region (Fig. 3). Figure 4 shows the solutions of Eq. (9) for different values of the doping density $N_2$. If the doping densities are equal ($N_1 = N_2$), then the spin polarization density decreases exponentially with distance from the injection point. The spin polarization density increases past the barrier, as the doping density $N_2$ increases.

Physically, the mechanism of the spin polarization density amplification near the boundary where doping is changed can be understood as follows. The spin polarization density can be increased near the boundary due to the charge localization: in the $N_2$ semiconductor region, the density of the electrons must be high. The electrons moving fast in the $N_1$ region, then move slowly in the $N_2$ region and gather in a small spatial region near the boundary.

**Conclusions:** Electron spin transport through the boundary between two semiconductor regions with different doping levels could lead to the electron spin polarization amplification near the boundary. The built-in electric field at the boundary accelerates propagation of the spin polarization through the boundary, if spin polarization passes from the low doping region to the high doping region. Spin amplification occurs past the boundary, within the distance of the order of the depletion layer width. It must be emphasized that there exists other mechanisms allowing increasing spin polarization density near the boundary. For example, the semiconductor regions could have different diffusion coefficients; for a more efficient spin focusing near the boundary, a lower diffusion coefficient of the $N_2$ region would be desirable. However, as mentioned in the introduction, the mechanism considered here, involving only the doping variation, has the advantage of not requiring a materials interface, thus avoiding additional spin-polarization losses.

**Acknowledgments:** We gratefully acknowledge helpful discussions with Profs. M.-C. Cheng and V. N. Gorshkov. This research was supported by the National Science Foundation, grants DMR-0121146 and ECS-0102500, and by the National Security Agency and Advanced Research and Development Activity under Army Research Office contract DAAD 19-02-1-0035.

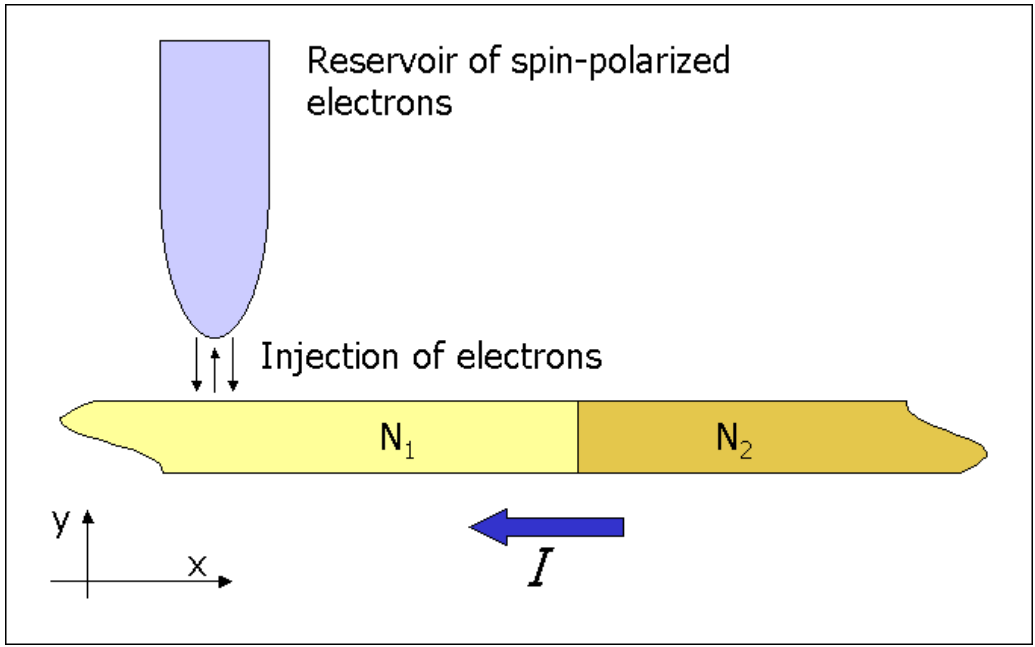

**Figure 1.** Injection of spin-polarized electrons in the system with two levels of doping.

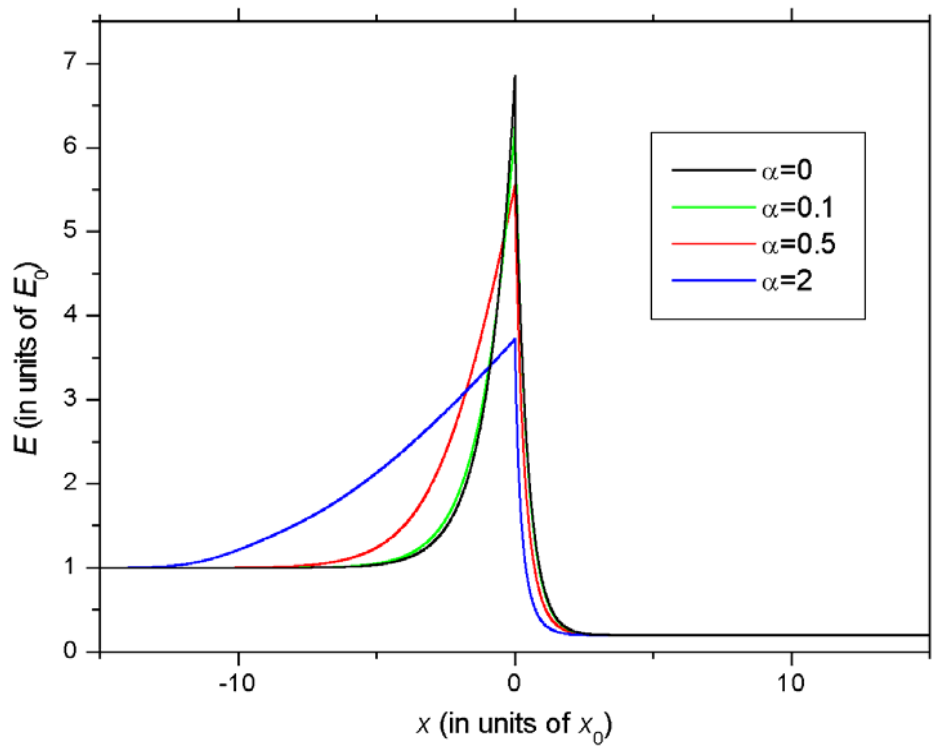

**Figure 2.** Electric field profile near the boundary, $N_2/N_1 = 5$, $\beta = -19$.

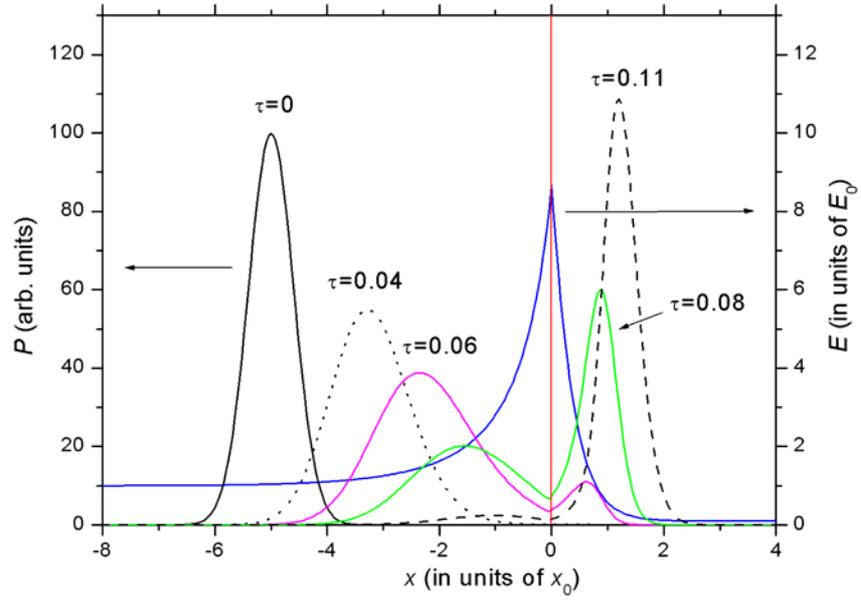

**Figure 3.** Dynamics of propagation through the boundary, of spin-polarized electrons injected at $\tau = 0$ ($\tau = t/\tau_{sf}$), $N_2/N_1 = 10$. The blue line denotes the electric field, the others lines show the distribution of the spin polarization density at different moments of time.

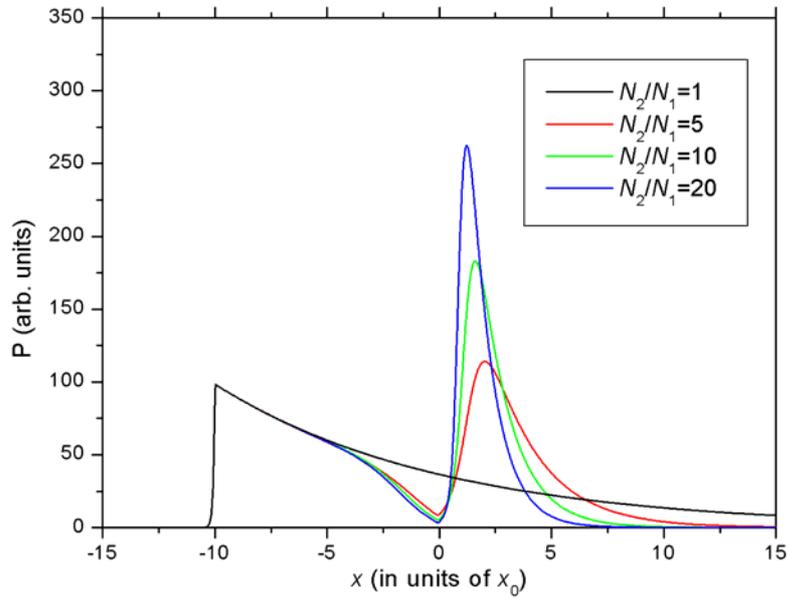

**Figure 4.** Distribution of the spin polarization density created by a point source located at $x = -10$. Spin accumulation effect near the boundary becomes more pronounced with increase of $N_2$.